# Autonomous Mobile Clinics: Empowering Affordable Anywhere Anytime Healthcare Access

Shaoshan Liu, Yuzhang Huang, Leiyu Shi

## Abstract:

Autonomous Mobile Clinics (AMCs) solve the healthcare access problem by bringing healthcare services to the patient by the order of the patient's fingertips. Nevertheless, to enable a universal autonomous mobile clinic network, a three-stage technical roadmap needs to be achieved: In stage one, we focus on solving the inequity challenge in the existing healthcare system by combining autonomous mobility and telemedicine. In stage two, we develop an AI doctor for primary care, which we foster from infancy to adulthood with clean healthcare data. With the AI doctor, we can solve the inefficiency problem. In stage three, after we have proven that the autonomous mobile clinic network can truly solve the target clinical use cases, we shall open up the platform for all medical verticals, thus enabling universal healthcare through this whole new system.

## 1. Introduction

We are facing a global crisis today as the healthcare cost is ever climbing, but with the aging population, government fiscal revenue is ever dropping [1]. This creates growing friction between different governments and their citizens. The people demand better and more affordable healthcare, but this becomes more and more unreachable with the current healthcare system.

Particularly, the United Nations Sustainable Development Goal 3 (SDG3) on health represents a universal recognition that health is fundamental to human capital and social and economic development. However, despite the progress achieved at the global level, many health systems are not sufficiently prepared to respond to the needs of the rapidly aging population. Inequitable access to healthcare continues to impede progress towards achieving universal health coverage. Therefore, provision of essential health services that's safe, affordable, and effective, with a special emphasis on the poor, vulnerable and marginalized segments of the population has become the main aim and task for all stakeholders in human development [2].

Under the current situation, there are only three options left for each government, higher taxes, worse healthcare services, or a more efficient healthcare system. From the perspective of human rights, *there is only one viable option, a more efficient healthcare system*.

According to a joint study by the World Bank and the World Health Organization, primary care is an effective way to reduce overall healthcare expenditure and, at the same time to improve healthcare service quality [3]. This theory has been proven in numerous developed countries, but access to healthcare, especially primary care, is difficult in many less developed countries. This problem further exacerbates in the current COVID-19 situation [4,5].

In this article, we introduce the concept of Autonomous Mobile Clinics (AMCs) with the vision to provide universal healthcare access to patients around the world [6,7,8]. We delve into the cause of deploying AMCs, the technologies enabling AMCs, the healthcare usage scenarios of AMCs, the potential benefits on healthcare management brought by AMCs, and health policy goals to integrate AMCs into the healthcare system. It is our hope that this comprehensive introduction of AMCs can inspire more research around innovative healthcare delivery methods for the benefits of humanity.

# 2. Grand Vision: Affordable Anywhere Anytime Healthcare Access Challenges

To create a more efficient healthcare system, we can start by enabling affordable anywhere anytime healthcare access through technological innovations, starting with primary care.  Three technical challenges immediately present themselves: ***The first challenge is access itself***; people get denied access to healthcare services for various reasons, including lack of mobility, particularly for disabled people and elderly, unpleasant user experiences, especially for countries with overloaded public healthcare systems, *etc.* We believe that an autonomous mobile clinic that comes to you by order of your fingertips solves the access problem.  We have successfully demonstrated the effectiveness of autonomous vehicles in addressing mobility problems, especially for the elderly and the disabled population [9].

***The second challenge is equity***; while mobility enables access, it does not guarantee equal access. The reality is that healthcare resources are unevenly distributed [10], and this distribution is often strongly correlated with wealth distribution. The equity problem again can be solved by technology; telemedicine on the mobile clinic connects the patient to medical experts worldwide [11], for the first time in human history, disrupting the uneven distribution barrier to truly enable equal health care access.  In our real-world autonomous vehicle deployments, we have encountered and provided a reliable solution to the connectivity problem [12].

***The third challenge is efficiency***; as the overall healthcare cost is climbing, people have high hopes for new technologies to improve healthcare access and quality while minimizing costs. We can consolidate clean health monitoring and diagnostic data through autonomous mobile clinics to train an AI doctor for the first time, which has been proven to be highly cost-efficient [13].  An AI doctor will handle most pre-screening and diagnostics tasks in primary care, freeing the human doctors for more sophisticated tasks.

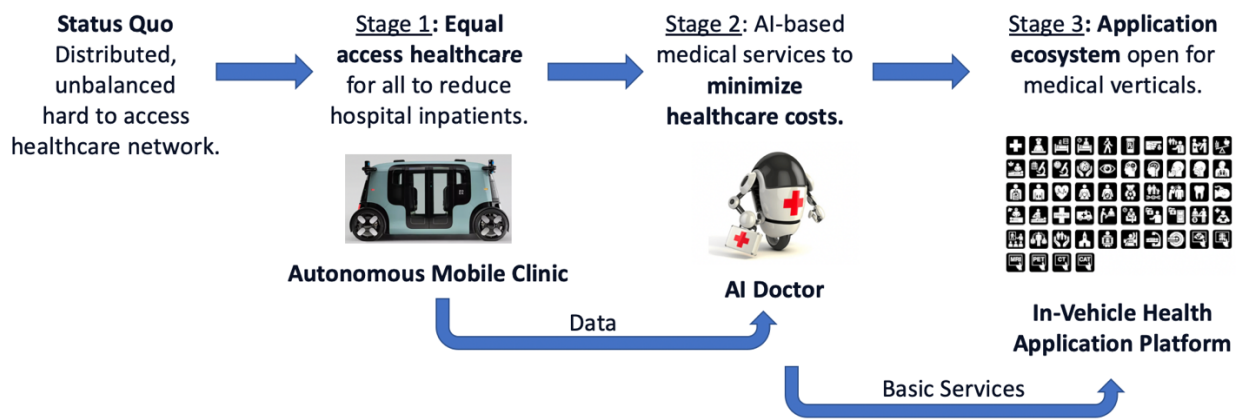

Figure 1: Development Stages of Autonomous Mobile Clinics

# 3. Autonomous Mobile Clinics: Enabling Technologies

In this section, we briefly introduce the enabling technologies enabling AMCs, a detailed technical overview can be found in [14]. Figure 2 illustrates how AMCs rely on multiple pillars of technical areas, including autonomous driving, telemedicine, medical equipment integration, and electronic health records:

- <u>Autonomous Driving</u>: level 4 autonomous driving is required for autonomous healthcare delivery. Each AMC needs to consist of a fully drive-by-wire chassis with fail-over backup systems, essential sensors for panoramic perception, a reliable localization system that can achieve decimeter level accuracy, a planning and control system that enables the vehicle to autonomously navigate through the target environment, as well as a High-Definition map of the operation area.

- <u>Telemedicine</u>: network connectivity is key to guarantee the Quality-of-Service (QoS) of telemedicine applications. there are two categories of data communication for in-vehicle healthcare services: healthcare sensing data and video data stream. Healthcare sensing data involves the transmission of raw sensing data collected from healthcare sensors, such as imaging devices, whereas video streams allow interactive communication between the patient and the remote doctor.
- <u>Medical Equipment Integration</u>: a mobile clinic requires the integration of many medical equipment. However, each medical equipment manufacturer enforces its own data standard, preventing the seamless healthcare delivery experience, especially that the equipment deployment needs to be highly customized to the specific healthcare service scenarios. A standard method of integrating and making sense of the enormous amount of sensing data is required.
- <u>Electronic Health Records</u>: electronic health records are key to enabling integrated care and to the training of AI doctors. With AMCs, for the first time we are able to collect integrated health data instead of having to collect fragmented health data from each department. This is the prerequisite for AMCs serving as a touch point of integrated care. In addition, by combining the integrated health data and diagnostics from the remote doctor, we can develop AI systems to handle basic diagnostic functions to reduce overall healthcare costs.

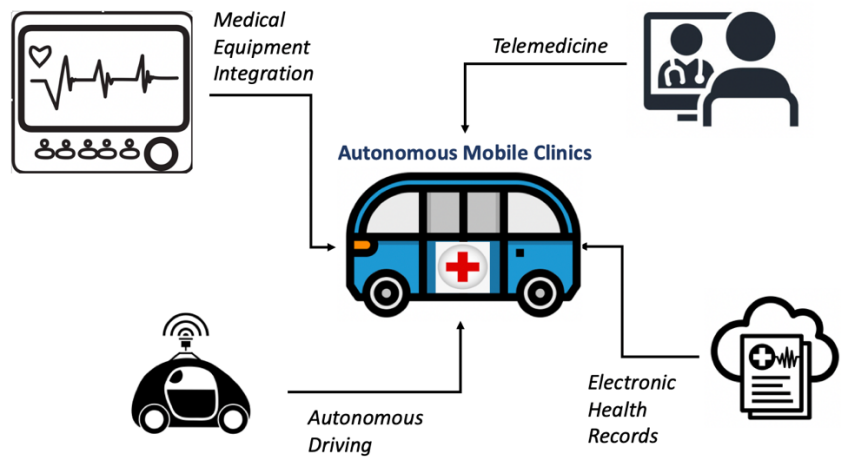

Figure 2: Enabling Technologies for Autonomous Mobile Clinics

## 3. Autonomous Mobile Clinics: Healthcare Usage Scenarios

To address the aforementioned challenges, we advocate that autonomous mobile clinics provide a revolutionary and effective way of healthcare service delivery. The idea of autonomous mobile clinics is not new, concept products such as Toyota e-Care [15] have emerged in recent years, but only stayed in the concept stage. As shown in Figure 1, to enable a universal autonomous mobile clinic network, a three-stage technical roadmap needs to be achieved: In stage one, we focus on solving the inequity challenge in the existing healthcare system by combining autonomous mobility, telemedicine, and primary care. In stage two, we develop an AI doctor for primary care, an AI doctor that we foster from infancy to adulthood with clean healthcare data. With the AI primary care doctor, we can solve the inefficiency problem. In stage three, after we have proven that the autonomous mobile clinic network can truly solve the primary care problem, we shall open up the platform for all medical verticals, thus enabling universal healthcare through this whole new system.

Autonomous mobile clinics also bring along multiple immediate clinical benefits. First, especially under the current COVID-19 situation, autonomous mobile clinics provide a perfect environment for Point of Care Test (POCT), which refers to multiple clinical tests carried out at the point of care near the patients, rather than from clinical laboratory. When dealing with patients of infectious diseases, lab tests such as White Blood Cell test and C-reactive Protein are of the utmost importance. Autonomous mobile clinics provide an isolated environment for POCT at the location where the patients reside, effectively constraining the spread of infectious diseases.

Under COVID-19, medical resources become even more scarce and many patients suffering from serious health conditions, such as Myocardial infarction, Stroke or Diabetes fail to get proper testing, such as creatine kinase myocardial band (CK-MB) and blood glucose, let alone treatments [16]. The multiplex POCT model includes regular gases of bloods, metabolites, as well as patient's electrolytes, and various immunoassays [17]. By following the multiplex POCT model to integrate the medical instruments in the autonomous mobile clinics, we can greatly reduce the lag time on transferring blood sample to laboratory as well as enhancing the diagnostic efficiency.

Taking Acute Myocardial Blood Marker Test as an example, it has been confirmed that POCT introduces a mean reduction of 110 minutes of test result turnaround time compared with traditional laboratory testing methods [18], the same benefit has been observed in the Emergency Department as well [19]. The second benefit of autonomous mobile clinics is the combination of the multiplex POCT model with telemedicine to carry out online healthcare consultation services to patients. With this capability, remote doctors can review patient's lab test results in real-time, hence the doctors are able to perform consecutive screening, monitoring, as well as to provide simple instructions on treatment process. These all take place on the point of care without having physical contact with the patients, greatly reducing the risks of contracting infectious diseases. We summarize the potential benefits of performing a few POCTs on autonomous mobile clinics as follows:

Table 1: Point of Care Test (POCT) Effectiveness of AMCs

| POCT on Autonomous Mobile Clinics | Benefits for the Patient | Benefits for the Healthcare Provider |
|---|---|---|
| Blood Creatinine | Reduction of 60 min of result turnaround time. | Increase of diagnosis and treatment efficiency. |
| Blood gas and electrolytes | Reduction of 120 min of vital analysis time, especially for emergency patients. | Enhanced rescue efficiency and potentially reduce mortality. |
| Blood Glucose | Reduction of 90 min of diagnosis time, especially on patients with diabetes co-morbidity, such as hypertension, renal disease, cerebral vascular disease. | Shorten time on accessibility of related medical resources; increase diagnosis precision; reduce mis-diagnosis on complications. |
| Telemedicine | Reduction of 120 min of waiting time. | Enhance early Diagnosis of acute and chronic conditions. |

# 4. Potential Benefits on Healthcare Management

From the New England Journal of Medicine report "Telemedicine Is Mainstream Care Delivery", the authors conclude that "Mobile + Medical care" is an essential way to improve healthcare efficiency, the report strongly advocates major hospitals to develop remote consultation systems, whether it is mobile or static, to address the imminent problem of healthcare access [20]. AMCs address this exact problem, and in this section, we summarize the potential benefits AMCs bring to healthcare management:

1. **Optimal medical resources allocation**: The AMC remote consultation system can break geographical boundaries, not only enable patients in different areas to access medical services, but also improve the level of medical services in poor population, so that medical resources can be allocated and used more reasonably [21]. The ultimate goal of the AMC project is to reach universal healthcare access. Unfortunately, today half the world's population lacks access to essential health services [22]. The per capita healthcare spending in the U.S. is $12,000, whereas the per capita healthcare spending in the least developed countries (LDCs) is merely $42. AMCs have the potential to narrow this tremendous resource allocation gap and enable a healthier population in LDCs, which is essential to boost the economy in LDCs. We are currently conducting a thorough study to understand the full impact of AMCs on the world economy.
2. **Affordable medical training**: One of the key roadblocks for universal healthcare access is the lack of healthcare professionals. Today, the global medical education market size is only around USD 30 Billion, and the average cost of training a medical doctor in the U.S. is around $200,000. This extremely high cost barrier for people in LDCs. AMCs can not only be used for disease screening, diagnosis, and treatment, but also for telemedicine training, education, and international telemedicine exchanges. From this perspective, with AMCs, hospitals across the world can form medical education alliance to develop medical training curriculum and to deliver medical training to medical trainees and healthcare professionals around the world through AMCs. Improving medical skills and leveling the healthcare service qualities across the globe is another key objective of the AMC project.
3. **Healthcare cost reduction**: On average, a primary care doctor can treat at most 20 patients per day, and the average annual salary of a primary care patient is around $200,000 in the U.S. This cost is unbearable for people in LDCs. In contrast, AI doctors within AMCs can treat unlimited patients with virtually no additional cost. Therefore, the AMC platform is able to minimize healthcare cost. Today, many AI doctors are already able to perform basic disease screening to increase the number of patients a primary care doctor can treat today. As a next step of development of AMC, we plan to focus on developing AI doctors for the top ten diseases in LDCs.

## 5. Health Policy Framework

A bold technical roadmap also exposes many legal and policy challenges, including autonomous vehicle safety, data security, data privacy, medical service delivery, and many more. Navigating through this regulatory matrix would be extremely challenging, as reported by many artificial intelligence projects [23]. In this section, we summarize the health policy framework to facilitate the integration of AMCs into our healthcare systems.

1. **Standard Operating Procedures**: AMCs are an innovative method of healthcare service delivery. To ensure the most appropriate and quality care for the patients, a set of standardized healthcare service delivery procedures, or Standard Operating Procedures (SOPs), need to be developed for AMCs [24]. Specifically, a SOP is defined as a written set of instructions that a healthcare professional should follow to complete a job safely, with no adverse effect on personal health or environment and in a manner that maximizes the probability of a beneficial health outcome in an efficient manner. In the case of AMCs, interactive technologies are required to guide the patients through the SOP process for them to properly receive healthcare service.

2. **Technical Standards**: AMCs consist of a large set of software and hardware, potentially from a large number of vendors. While after decades of development, the industry has gradually conformed to a set of standards for telemedicine, technical standards for integrating a varying set of hardware and software for AMCs are still in their infancy, and these need to be defined clearly for AMCs to reach large-scale deployments.

3. **Information Privacy and Security**: similar to technical standards, information privacy and security standards have gradually matured after years of development, especially during COVID-19 [25]. AMCs may expose new security and privacy problems as many disease screening and triage tasks are carried out by the AI doctors onboard of AMCs. Many governments may restrict the usage of AMCs due to the lack of regulatory measures and standards to enforce information privacy and security.

4. **Effectiveness Evaluation Metrics**: to aid policy makers to make decisions on whether to adopt AMCs into their healthcare systems, and on how and where to deploy AMCs, a set of commonly agreed on effectiveness evaluation metrics need to be developed to ensure fair comparison. Similar studies have been done for mobile care, for instance, a public health study demonstrated that mobile care effectively increases acute care efficiency by over 36%, providing strong evidence favoring the utilization of mobile care [26].

5. **Education and Training**: proliferation of AMCs will create a whole new category of healthcare jobs, including technicians certified to work on AMC equipment, nurses or healthcare professionals onboard AMCs to help patients, or remote doctors giving medical treatment based on diagnostic data collected on AMCs. Training and certifying this new category of AMC healthcare professionals imposes a new challenge for the existing education and training system. Fortunately, this problem is not unique to AMC, but very common in the emerging fields such as Artificial Intelligence, and many learnings from Artificial Intelligence can be applied to AMCs [27].

6. **Institutional Change**: AMCs require integration of expertise from many different traditional areas, and today each of these areas is regulated separately. Take the U.S. for example, the National Highway Traffic Safety Administration (NHTSA) regulates autonomous vehicle safety, the Federal Trade Commission (FTC) regulates data security and privacy, and the Food and Drug Administration (FDA) regulates medical service delivery. Besides regulations, standard bodies like IEEE and ISO develop standards for applications of emerging technologies, such as the autonomous mobile clinics. Hence, the emergence of AMCs demands an institutional change to form a unified regulatory body for all matters related to AMCs.

# 6. Summary

Autonomous Mobile Clinics (AMCs) have the potential to solve the healthcare access problem by bringing healthcare services to the patient by the order of the patient's fingertips. In this article, we have introduced the vision of AMCs, the enabling technologies for AMCs, the healthcare usage scenarios of AMCs, the potential benefits on healthcare management brought by AMCs, and health policy goals to integrate AMCs into the healthcare system. It is our hope that this comprehensive introduction of AMCs can inspire more research around innovative healthcare delivery methods for the benefits of humanity.